\documentclass[preprint,3p,times,twocolumn]{elsarticle}
\journal{Journal of \LaTeX\ Templates}

\usepackage{lineno,hyperref}
\modulolinenumbers[5]

\journal{Journal of Computer-Aided Design }

\usepackage{times}
\usepackage{epsfig}
\usepackage{graphicx}
\usepackage{amsmath}
\usepackage{amssymb}
 \usepackage{subfig}
\usepackage{slashbox}
 \usepackage{xcolor}
\usepackage{algorithm}  
\usepackage{algorithmic}  
\usepackage{egweblnk}
\usepackage{bm}
\usepackage{array}
\usepackage{mathrsfs}
\usepackage[numbers]{natbib}
\usepackage{multicol}
\usepackage{multirow} 
\usepackage{bm}
\usepackage{booktabs}
 \usepackage{color}
\definecolor{junz}{rgb}{0,0,0}

\begin{document}
\begin{frontmatter}

\title{Normal Estimation for 3D Point Clouds via Local Plane Constraint and Multi-scale Selection}

\address[1]{School of Mathematical Sciences, Dalian University of Technology, Dalian, 116024, China}
\cortext[mycorrespondingauthor]{Corresponding author}
\author[1]{Jun Zhou \corref{mycorrespondingauthor}}
\ead{zj.9004@gmail.com}
\author[1]{Hua Huang}
\author[1]{Bin Liu}
\author[1]{Xiuping Liu}


%
%


\begin{abstract}
In this paper, we propose a normal estimation method for unstructured 3D point clouds. In this method, a feature constraint mechanism called Local Plane Features Constraint (LPFC) is used and then a multi-scale selection strategy is introduced. The LPEC can be used in a single-scale point network architecture for a more stable normal estimation of the unstructured 3D point clouds. In particular,  it can partly overcome the influence of noise on a large sampling scale compared to the other methods which only use regression loss for normal estimation. For more details,  a subnetwork is built after point-wise features extracted layers of the network and it gives more constraints to each point of the local patch via a binary classifier in the end. Then we use multi-task optimization to train the normal estimation and local plane classification tasks simultaneously. Via LPFC, the normal estimation network could obtain more distinguish point-wise plane-aware features that can describe the differences of each point on the local patch. Finally, thanks to the distinguish features constraint, we can obtain a more robust and meaningful global feature that can be used to regress the normal of the local patch. Also, to integrate the advantages of multi-scale results, a scale selection strategy is adopted, which is a data-driven approach for selecting the optimal scale around each point and encourages subnetwork specialization. Specifically, we employed a subnetwork called Scale Estimation Network to extract scale weight information from multi-scale features. More analysis is given about the relations between noise levels, local boundary, and scales in the experiment. These relationships can be a better guide to choosing particular scales for a particular model.  Besides, the experimental result shows that our network can distinguish the points on the fitting plane accurately and this can be used to guide the normal estimation and our multi-scale method can improve the results well. Compared to some state-of-the-art surface normal estimators, our method is robust to noise and can achieve competitive results.
\end{abstract}


\end{frontmatter}

\section{Introduction}
Recently, due to the advances in 3D acquisition technologies,  point-based representations of complex objects and environments cloud be captured  in many applications such as the field of autonomous driving and robot manipulation. With the proliferate available data, the analysis of raw 3D point clouds has been paid great attention, and a significant problem in shape analysis is to robustly estimate the normals from a raw, unordered point cloud, aimed to handle the challenging difficulties of sampling density, noise, outliers. The precise normal estimation can be used as extra information for improving some computer vision tasks, such as reconstruction \cite{M2013Poisson}, registration\cite{Pomerleau2015A} and object segmentation\cite{Grilli2017a} problems.

The common approach is to sample local neighbors around each point and to estimate a point cloud normal based on the local points statistics by the fitting local surface. However, the robust computation of normals is always affected by many issues. Due to lack of connectivity or structure information, raw point clouds normal estimation always faces challenges, especially when the noisy, incomplete, and typically exhibit varying sampling density occur in real scan models.  The classic methods are often sensitive to the scale selection and unstable to the real scan noises. Moreover, many man-made objects exhibit sharp features like corners or edges, which are easily lost.  Although some learning-based methods are proposed in recent years, this kind of method is known to often achieve far better results compared to data-independent methods. However, deep learning methods roughly select local points as input and just train a model to regress the normal of each point commonly. They do not consider the outliers and the error points on the patches due to the different sample scale, either. Here we define the error points as the points that may not belong to the local fitting plane. In particular, with the increase of sampling radius or point cloud noise, the accuracy of deep learning methods are increasingly influenced by input. The SOTA methods proposed always do not consider that the input patches may include lots of error points. Intuitively, for a normal estimation network, if we have a stronger constraint on point-wise features of the input patch, this means that the constraint can better distinguish the point-wise features belonging to the fitting plane or non-plane points, we would obtain more accurate normals in any sampling scale.

Our key insight is that local shape properties can be robustly estimated by suitably accounting for shape features, noise margin, and sampling distributions. However, such a relation is complex and difficult to manually account for.  Hence, we propose a data-driven approach based on local point neighborhoods similar to PCPNET\cite{guerrero2018pcpnet}, which trains a point network to directly learn local properties using ground truth reference results under different input perturbations. But different from their architecture we consider producing a stronger constraint to the network to ensure that the more meaningful and plane-aware features can be obtained and we also use a scale selection method to ensure the right scale can be achieved. In this paper, we first proposed a Local Plane Features Constraint (LPFC) to better distinguish the features differences between these two kinds of points in each patch, and this plane-aware features enhancement item can improve the results of normal estimation. Especially, when the large noise occurs and for the regularity objects, our method can give an impressive estimation result. Also, our multi-scale strategy can guide the network to select the best scale for point estimation. By scale weight selected, our method can obtain more accuracy normal results. 

The main contributions of this paper are:
\begin{itemize} 
\item For single-scale normal estimation, Local Plane Features Constraint (LPFC) is used in our networks to ensure that the normal estimation network more robust to the noise point cloud in any sampling scale. Besides, the binary classifier used in our LPFC can well obtain the main part of the patch which is almost in the fitting plane and distinguish the error points, especially when the sampling scale is large.
\item A scale selection strategy for scale prediction is employed in our method. In this paper, we propose a novel scale estimation network, which is used to select the most suitable scale of each point through a joint analysis of multi-scale features extracted from single-scale networks.
\item  The experiment shows outperformance results in single-scale and multi-scale compared some state-of-the-art surface normal estimators.
\end{itemize}

\section{Related Work}
Normal estimation has a very long history in geometry processing, motivated in large part by its direct utility in shape reconstruction.  In this section, we present an overview of traditional normal estimation methods and learning-based methods for normal estimation. Here, we first  give a brief description of the 3D point deep learning  history.

\subsection{Deep learning for 3D point clouds}
The point cloud representation is challenging for deep learning methods because it is both unstructured and point-wise unordered. Early, the voxel-based shape representations \cite{maturana2015voxnet, wu20153d, qi2016volumetric} are adopted to direct extension of 2D image grids to 3D grids, which are the pioneers applying 3D convolutional neural networks. However, the volumetric representation is constrained by its resolution, and the local geometric information maybe lost. So this leaning-based representation methods can not be used to predict local property well. Recently, there has been quite a few developments in feature learning from 3D point cloud directly. The pioneers of the point cloud networks are PointNet\cite{qi2017pointnet}, PointNet++\cite{qi2017pointnet++} and PointCNN\cite{li2018pointcnn}. All the above-mentioned methods are used to learn global features for 3D object detection, classification and retrieval tasks.  Specially, The PointNet applies a symmetric, order-insensitive, function on a high-dimensional representation of individual points. This network can be used to extract geometric features directly from the original data. In this paper, we also employ the a architecture similar to PointNet \cite{qi2017pointnet} to directly extract plane-aware features  of the local points and to regress the normals of the 3D model in the end.

 \begin{figure*}
  \centering
    \includegraphics[scale=0.33]{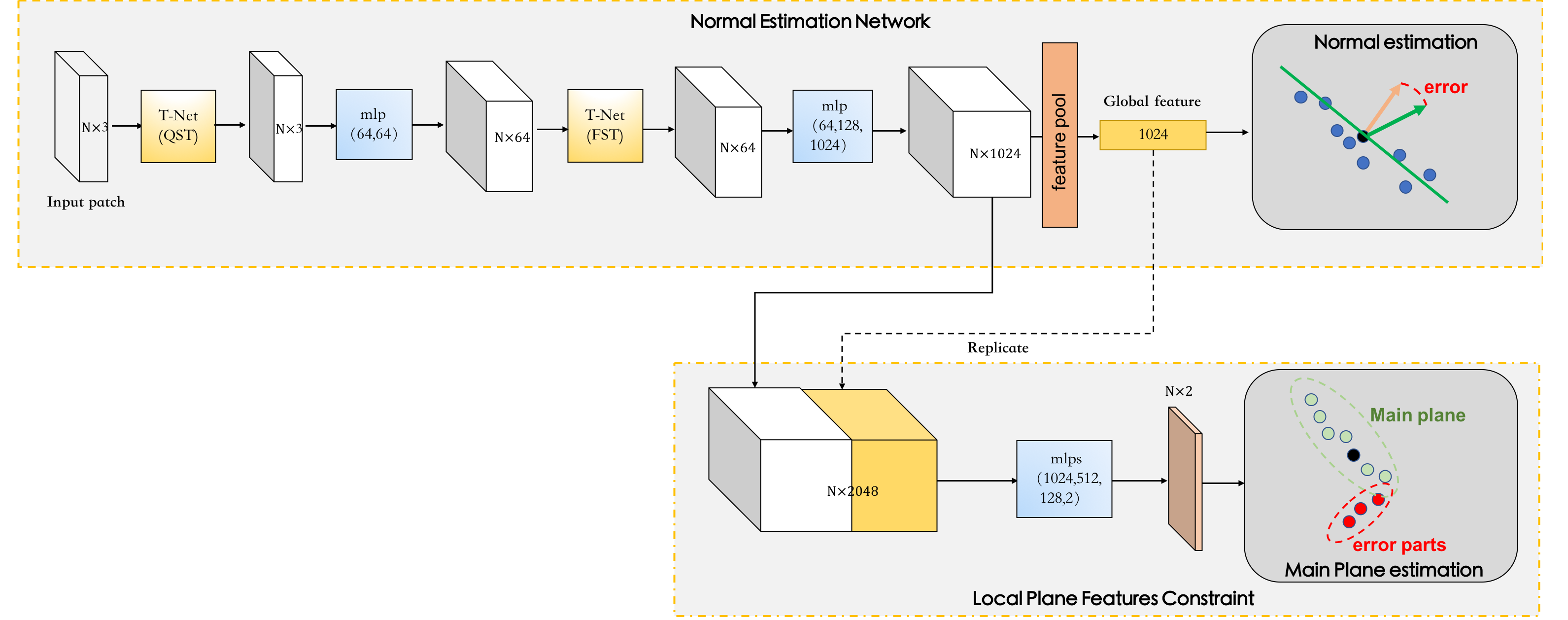}
  \caption{Our proposed network architecture for a single scale normal estimation. The input patch first goes through a feed-forward neural network to compute a 1024-dimension feature vector for each point. Here it splits into to branches: one for normal estimation and the other for main plane estimation.}
  \label{FIG:1}
\end{figure*}

\subsection{ Traditional normal estimation}
Early works on normal estimation are based on Principal Component Analysis (PCA) \cite{hoppe1992surface} and the variants.  In their work, the PCA method is used to solve the eigenvectors of  the covariance matrix constructed from the local patch points. In addition, singular value decomposition (SVD)\cite{klasing2009comparison} is also used for normal estimation. They both  specify the neighbors within some scale, and then uses PCA or SVD method to estimate a tangent plane.  The performance of these approaches usually dependent on the scale of the selected patches and the noise, outliers may also have some influence for the normal estimation. An obvious observation is that if the sampled patch contains boundaries or consisted of one more fitting planes, the final normal estimation will  get worse. Several methods have been proposed to address these limitations by both designing more robust estimation procedures, capable of handling more challenging data, and by proposing techniques for estimating normal orientation.  Then assigning Gaussian weights to the neighbors method\cite{Pauly2003Shape} or adapting the radius sampling method\cite{Mitra2004ESTIMATING} are proposed to  solve these common issues. Then Yoon et al. \cite{Yoon2007Surface} consider to assembling statistics techniques to improve the classic PCA method, then a more stable results can be estimated. Besides, some other approaches \cite{alexa2001point, cazals2005estimating, guennebaud2007algebraic} fitting higher-level surfaces like local spherical and quadrics surfaces.  These methods usually choose a large-scale neighborhood, leading them to smooth sharp features and also to fail in estimating normals near edges  and do not address another major issue of real-world point clouds. Another kind of approaches mainly relies on using Voronoi cells of point clouds\cite{Amenta1998Surface,Merigot2011Voronoi,Dey2004Provable}. Although these methods can improve the normal estimation on the sharp points, the estimation results are sensitive to the noise. To handle this difficult problem, Alliez et al. \cite{Alliez2007Voronoi} proposed a PCA-Voronoi method, which provides some control smoothness by grouping adjacent cells on the 3D model. 

While many of these methods hold theoretical guarantees on approximation and robustness, in practice all above methods require a careful setting of parameters, and often depend on special treatment in the presence of strong or structured noise. Unfortunately, there is no universal parameter set that would work for all settings and shape types. So the data-driven methods are urgently necessary technologies for more robust to estimate the normals of 3D models.
  
 \subsection{Learning-based normal estimation}
 Deep learning based approaches also found their way into surface normal estimation with the recent success of deep learning in a wide range of domains. Boulch  et al. \cite{Boulch2016Deep} are the pioneers to apply  Convolutional Neural Network to regress the point normal, they proposed to use a CNN on Hough transformed point clouds in order to find surface planes of the point cloud in Hough space. But this method is based on image input and do not use the point data directly. Recently, due to the advent of graph neural networks and geometric deep learning\cite{Bronstein2017Geometric}, then Charles et al. \cite{qi2017pointnet} proposed PointNet to directly learn features from points data. Inspired by PointNet, Guerrero et al. \cite{guerrero2018pcpnet} proposed a deep multi-scale architecture for surface normal estimation, the different from PointNet, they use patch points data as input and shows that a Quaternion  transformation is more useful to the normal estimation task and also show that mean pooling feature is better. However, their method do not do well in the scale selection due to simply concat multi-scale features of the network.  Later, Ben-Shabat et al. \cite{ben2019nesti} proposed to  use a mixture- of-experts architecture, which relies on a data- driven approach for selecting the optimal scale around each point and encourages sub-network specialization. However, due to use fisher vectors as input representation, their network can not obtain well results in a single scale, more information maybe lose in the process of computing point statistics only on a coarse Gaussian grid.
In addition, we find that even though the results of the final normal estimation can be improved by multi-scale joint decision making, the trained network itself cannot obtain the best results in any sampled scale.  The single-scale sampled patch itself  always has error points as described above. All above proposed methods   do not  consider this common situation in the stage of neighbor sampling. In this paper, we use a Local Plane Feature Constraint to obtain a plane-aware point-wise features in a patch, so the final global features are more robust to the error points and we also can used the subnetwork to obtain a more clean patch which only contain points on the main plane. Finally we also employ a multi-scale selection method to adaptive chose the best scale on each point.

\begin{figure*}
  \centering
    \includegraphics[scale=0.43]{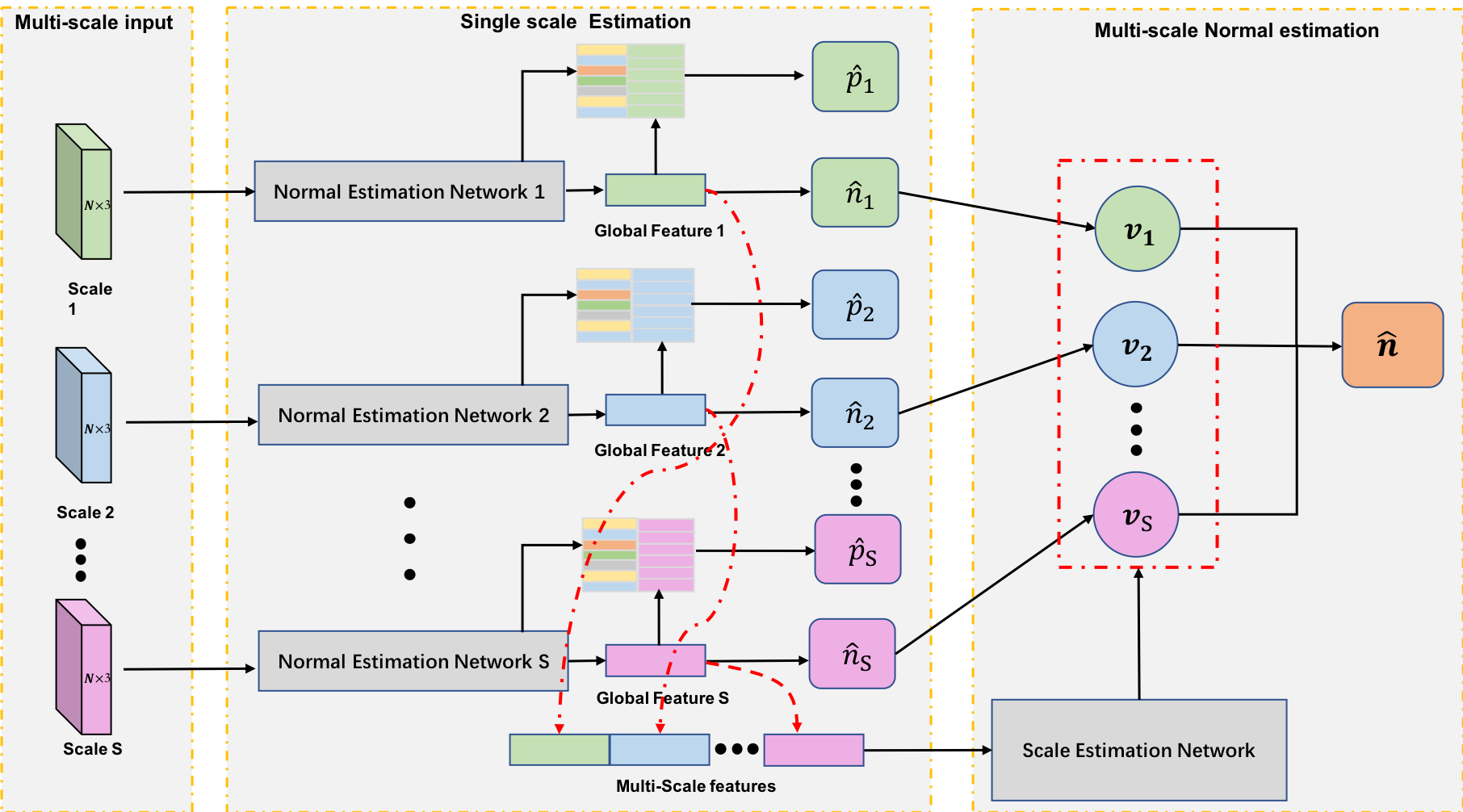}
  \caption{Multi-scale normals estimation architecture. Here, The S scales are used as input for our multi-scale networks. Then the normal of each scale is estimated based on the single scale network, and the global features are obtained via concatenated the features from each scale network. Finally, a Scale Estimation Network is used to determine the optimal scale and our network uses the corresponding single-scale sub-network to estimate the normal via learning scale weights $v_{1},v_{2},\cdots,v_{S}$.}
\label{FIG:2}
\end{figure*}

\section{Approach}
Our goal in this work is to estimate point normals from a point cloud utilizing features constraint that enhances regularity and scale selection strategy.  Similar to PCPNet\cite{guerrero2018pcpnet}, given a 3D point cloud $X =\{p_{t}|t=1, 2, \cdots,  N\}$, the input of our network are the local patches of this point cloud $X$, centered at points $p_{t}$ with a fixed radius (scale) $r$ proportional to the point cloud's bounding box extent, namely, $P_{t}^{r} =\{p|\|p-p_{t}\|_{2}<r, p \in X\}_{t=1,2,\cdots N}$. For the single scale network, our architecture consists of two main parts. First, given sampled neighbors with  fixed scale, we mainly learn plane-aware features on each PCPNet\cite{guerrero2018pcpnet} is that we proposed a new branch in our network to ensure that our network can extract the plane-aware features. Finally, a multi-task loss is used to obtain an important region of input and estimate the local normal meanwhile.  For multi-scale input, we also consider selecting the most appropriate scale for each point by employing a scale selection network. Figure \ref{FIG:1} and Figure \ref{FIG:2} show the single-scale feature constraint and multi-scale selection networks respectively.  More details are described in the following subsections.

\subsection{Pre-processing}
Similar to PCPNet \cite{guerrero2018pcpnet}, for each sampled local patch at the centered point $p_{t}$, we should first translate the patch into a local frame relative to the centroid point and normalize its size to one scale: 
 $p_{t}^{(j)} = (p_{t}^{(j)}-p_{t})/r $ for $t=1,2, \cdots, N $ and  $j=1, 2, \cdots, 500$, where $p_{t}$ is the coordinate of the centroid of patch $t$, and $r$ is the fixed scale radius. Our network takes a fixed number of points as input. Patches that have too few points are padded with replicate points of the patch, we think that process can keep the balance of points better compared to just padding zeros points, and we pick a random subset for patches with too many points.

\subsection{Plane-aware Features Constraint}
Given an input patch  $P_{t}^{r}$ centered at point $p_{t}$, different from the previous methods that regress the local normal directly. We consider applying a features constraint strategy for extracting plane-aware features to estimate the point normal more stably. During the training stage, we know that each patch  $P_{t}^{r}$ has a real normal $n_{t}$ at center point $p_{t}$ as ground truth,  the goal is to train a network that can regress the predict normal $\hat{n}_{t}$ to the real normals as close as possible. In our work, we consider to give stronger constrains to the normal estimation network at the point-wise level instead of at the patch level. To be specific, for training, we need to obtain  real normal not only  at the center point but also at other points in the neighbors. The real normals on the local patch are denoted  as $ N_{t} = \{n_{t}^{j}|j=1, 2, \cdots, K\}$, where $K=500$ is the number of points in a patch.  We then calculate the $L_{2}$ distance between each point normal to the center normal,  we use this distance to measure the error distance that each point to the real plane. 
\begin{equation}
P(n_{t}^{j}) = \min (\|n_{t}^{j}-n_{t}\|_{2},\|n_{t}^{j}+n_{t}\|_{2})
\end{equation}
here $n_{t}$ denotes the real normal of the center point and the length of $P$ is $K$. Intuitively, on one local point cloud, if the normal  on one point is consistent with the central point, generally we assume that the point is on the fitting plane and such a point is more important.  Then we can normalize the error value of each patch as:
\begin{equation}
P =\frac{P-\min(P)}{\max(P)-\min(P)+0.01}
\end{equation}

\begin{figure*}
  \centering
    \includegraphics[scale=0.54]{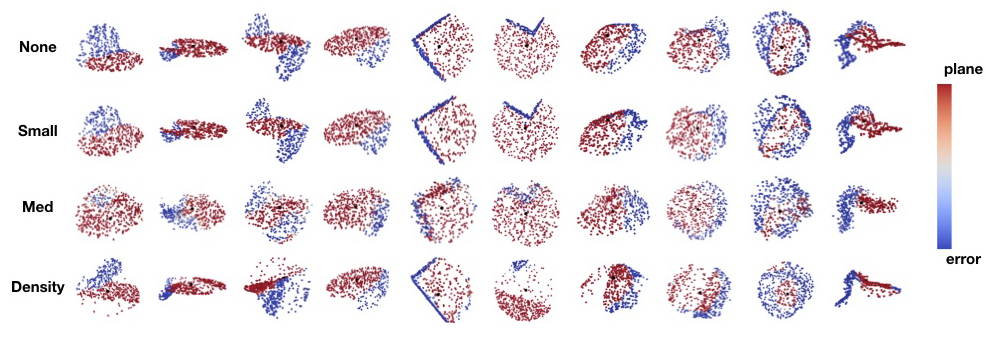}
  \caption{The local constraints plane produced by our LPFC. The proposed loss can give more constraints to the point-wise feature learning and an main plane region (red parts) is achieved through our network in different noise levels, which can have significant effects on the process of the normal estimation.}
  \label{FIG:03}
\end{figure*}

We use the threshold $\theta =0.5$ as the default value to choose a  main part in each patch as the ground truth $Y_{t} = [0,1]^{K}$ and the the Main Plane Estimation can be consider as a classification problem to predict the plane denoted as $\hat{Y}_{t} = [0,1]^{K}$.  We define the ground truth of the plane as:

\begin{equation}
Y_{t}(j) =
 \left \{  
 \begin{aligned}
 &1, & if ~~ P(n_{t}^{j}) \leq \theta\\
 &0,   & if ~~ P(n_{t}^{j}) > \theta
 \end{aligned}
\right.
\end{equation}
here $Y_{t}(j)=1$ means that the point $p_{t}^{(j)}$ is a plane point. For small scale, we need change this threshold value to weaken the plane constraint, and we use 0.8 for small scales (0.03). We consider learning point-wise features by classifying the input patch points as two parts, namely the main plane part and error part. Constraining the features learning process in our network, we can obtain more meaningful  features  to regress the normal of one patch. This binary classification constraints can improve the point network to extract regular features contained more plane information of the patch.  We call this constraint strategy as Local Plane Features Constraint (LPFC). Figure\ref{FIG:03} shows some estimation results via our network and we can see that the constraint is meaningful for the normal estimation.

\subsection{Single Scale Network}
Our  network follows the PointNet architecture and this is also similar to PCPNet\cite{guerrero2018pcpnet} but we also propose a new plane estimation subnetwork to constrain the point-wise features learned, then the plane-aware features are aggregated to global features of the local patch to regress the local normal.  An overview of the architecture is shown in Figure \ref{FIG:1}. 

\textbf{Local Plane Features Constraint.} In the part of point-wise features constraint network, we also use some layers of perceptrons to a classifier. The point-wise features can be constrained, some points implicit main plane information and some others have the error plane information. Through the back propagation optimization algorithm,  the features of each point may contain more information related to the normal fitting plane. The subnetwork is also shown in Figure \ref{FIG:1}.

\textbf{Normal Estimation.} For the normal estimation part of the network, we also use Quaternion spatial transformer to transfer the input patch to a canonical  pose. Here a subnetwork is used to evaluate the quaternion that parameters a local  spatial transform. The output of Quaternion can be converted to a rotation matrix.  Then the rotated patch is feed into the normal estimation network. One important property of the network is that it should be invariant to the input point ordering. Qi et al.\cite{qi2017pointnet} show that this can be achieved by applying a set of functions with shared parameters to each point separately and then combine the resulting values for each point using a symmetric operation. In our work, we use the weight mean pooling function to obtain the patch feature provides a rich description of the patch, and the global feature is also used in the main plane estimation to constrain the global and point-wise features. We use a three-layer fully connected network to perform the normal regression. As shown in the top part of Figure \ref{FIG:1}.

\textbf{The Multi-task Loss.} The network then diverges into two different branches performing two tasks: predicting the center normal of the input patch and constraining their point-wise features to better select the main part in one patch. The loss of our network is the sum of the losses of its two branches,
\begin{equation}
\mathcal{L}_{total}= \mathcal{L}_{normal}+\mathcal{L}_{main}
\end{equation}
The normal prediction loss performed by minimizing the Euclidean distance between estimated normals $\hat{n}$ and ground truth normals $n$  as usual. 
\begin{equation}
\mathcal{L}_{normal}=\frac{1}{N}\sum_{i=1}^{N}\min(\|n_{i}-\hat{n}_{i}\|_{2},\|n_{i}+\hat{n}_{i}\|_{2})
\end{equation}
here $N$ is the total number of the point cloud. We also employ a discriminative function to present constraint the patch point-wise features learning and this loss can improve the robustness of the network for both point-wise feature learning and also obtain an effective global feature to estimate the final normal. They can be seen as  a binary classification problem which can distinguish the fitting plane points and other error points in one patch and this loss is defined by the cross-entropy loss :
\begin{equation}
\mathcal{L}_{main}=\frac{1}{N}\sum_{t=1}^{N}\sum_{y_{i}\in Y_{t}}y_{i}\log(\hat{y}_{i})+(1-y_{i})\log(1-\hat{y}_{i})
\end{equation}

 \begin{figure}[h]
  \centering
    \includegraphics[scale=0.32]{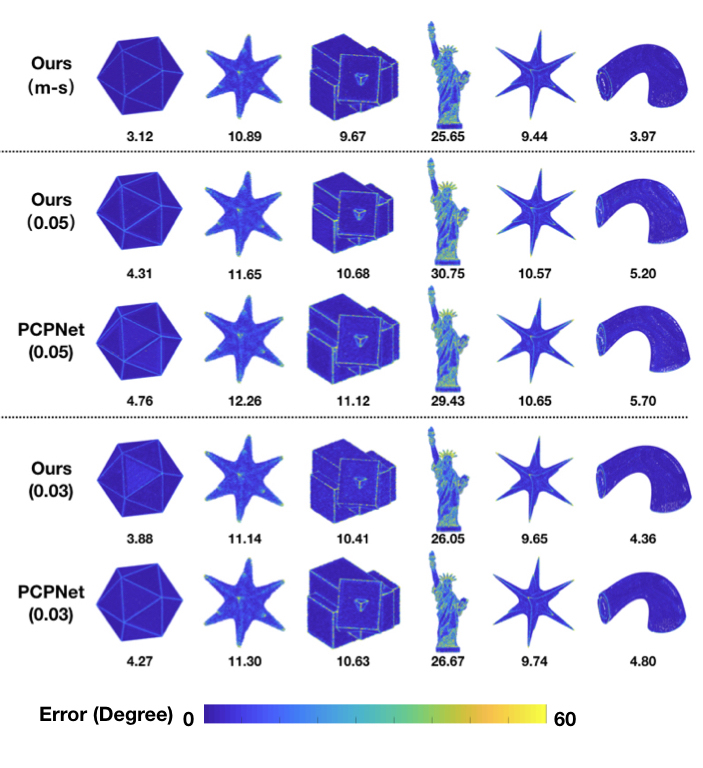}
  \caption{Comparison of the angle error for normal estimated of our method using multi-scale (m-s) with scale selection and single scales (0.03 and 0.05) with the LPFC to PCPNet\cite{guerrero2018pcpnet}.}
  \label{FIG:05}
\end{figure}

\subsection{Multi-Scale Normal Estimation}
As shown in Figure \ref{FIG:2}. We give the pipeline of our multi-scale architecture to adaptively select best normal from $S$ single scale results. In our method, the normal is finally estimated using $S$ separate normal estimation networks. Each is a normal estimation network that has been introduced in the previous section, the single scale network estimates the patch normal at the center point and also estimates the planer points of the patch. The multi-scale adaptive selection network uses $S$ subnetworks from small scale to large scale. As shown in the third  part of Figure \ref{FIG:2}, each scale network outputs a three-element vector $\hat{n}_{s} = (\hat{n}_{x} , \hat{n}_{y} , \hat{n}_{z} )_{s}, s=1, 2,\cdots, S$.  The plane estimation constraints are also used in each scale and $\hat{p}_{1}, \hat{p}_{2}, \cdots, \hat{p}_{S}$ are given. The global features obtained from subnetworks are concatenated as one vector and this vector are feed to the Scale Estimation Network to obtain the weights of the S scale normals denoted as $v_{1},v_{2},\cdots, v_{S}$. The final predicted normal (for center point) is $\hat{n} = \hat{n}_{argmax (v_{s})}$, i.e. the normal associated with the subnetwork expected to give the best results. For the multi-scale network, we also train the network to minimize the difference between a predicted normal $\hat{n}$ and a ground truth normal $n$.  Finally, we minimize the total loss:
\begin{equation}
\mathcal{L}_{multi} = \sum_{s=1}^{S}v_{s}\cdot \mathcal{L}_{normal}^{s}+\sum_{s=1}^{S}\frac{1}{S}\mathcal{L}_{main}^{s}
\end{equation}
here, $\mathcal{L}_{normal}^{s}$ is the loss used for normal estimation in single scale $s$. Using this loss, each scale normal estimation network is rewarded for specializing in a specific input type. Note that during training, all the scale normal vectors are predicted and used to compute the loss and derivatives. However, at test time, we compute only one normal, which is associated with the maximal $v_{s}$.

\section{ Evaluation and Discussion}
\subsection{PCPNet dataset}
Our method is trained and validated quantitatively on the PCPNet dataset as provided by Guerrero et al. \cite{guerrero2018pcpnet}. It contains a mixture of point clouds sampled from man-made objects and high-resolution scan models such as bunny. Each point cloud consists of 100k points. We reproduce the experimental setup of \cite{guerrero2018pcpnet,ben2019nesti}, training on the provided split containing 32 point clouds under different levels of noise. The test set consists of six categories models which contain four sets with no noise, small noise ($\sigma = 0.00125$), medium noise ($\sigma =0.0065$) and high noise ($\sigma = 0.012$)  and two sets with different sampling density (striped pattern and gradient pattern).  Then, the Root Mean Squared Error (RMSE) on the provided test set is used as performance metric following the related works, where the RMSE is first computed for each test point cloud before the results are averaged over all point clouds in one category. 

\subsection{Training details}
The variants of our network are trained by using all patches sampled from 32 shapes, which includes 8 point cloud models with 4 levels of gaussian noises. For each sample point, we randomly extract 500 neighboring points enclosed within a sphere of radius. We train our single-scale network with a patch size of 0.01, 0.03 and 0.05 respectively. Finally, we also use a learned multi-scale selection approach to refine the normal results based on the 
multiple scales' results together. If the neighborhoods within the radius have more than 500 points, we perform random sampling from the local subset points and for those with fewer points within the patch radius, we just repeatedly and uniformly sample points from the local subset. In addition, we train our network by used Pytortch\cite{paszke2017automatic} on a single 1080 ti GPU.

Similar to PCPNet\cite{guerrero2018pcpnet}, we train our networks (single-scale and multi-scale) for up to 2000 epochs until convergence on the PCPNet dataset. A full randomization of the dataset, mixing patches of different shapes in each batch, was vital to achieve stable convergence. All our training was performed by using stochastic gradient descent with batch size 64 in the single scale training stage and the batch size is set to 16 in the multi-scale training stage. The learning rate is  $10^{-4} $ and momentum is set to $0.9$. In the stage of evaluation, we estimate the unoriented normals of all points on the models and the error compared to ground truth are calculated for evaluation.

\begin{table*}[]
\centering
\caption{The quantitative comparison of the RMS angle error for normal vector estimation of our method to other deep learning methods (Nesti-Net\cite{ben2019nesti}, HoughCNN\cite{Boulch2016Deep} and PCPNet\cite{guerrero2018pcpnet}) on a single scale.  The LPFC can improve the normal estimation when the noises occur on the shapes.}

\label{Tab01}
\begin{tabular}{lccccccccc}
\toprule
\multirow{2}{*}{} & \multicolumn{3}{c}{\textbf{Our Method}} & \multicolumn{3}{c}{\textbf{PCPNet\cite{guerrero2018pcpnet}}} & \multicolumn{2}{c}{\textbf{Nesti-Net\cite{ben2019nesti}}}&\textbf{HoughCNN\cite{Boulch2016Deep}}\\
\cmidrule(r){2-4} \cmidrule(r){5-7} \cmidrule(r){8-9} \cmidrule(r){10-10}
&  $0.01$      &  $0.03$   &   $0.05$
&  $0.01$      &  $0.03$   &   $0.05$
&  $0.01$      &   $0.05$  
&  $0.05$ \\
\midrule
No Noise  &\textbf{8.47}    &\textbf{8.67}    &9.81   &8.51   &8.69   &9.68   &9.32         & 12.73   & 12.73          \\
\hline
\textbf{Noise} \\
Small Noise   &11.6                         &\textbf{10.49}                   & 11.47                   & 11.49          & 10.61         &\textbf{11.46}         &\textbf{11.31}      & 13.36           & 11.62                  \\
Middle Noise           &\textbf{35.69}                         & \textbf{17.62 }                   & \textbf{17.93 }                  & 36.39           & 18.02          & 18.26          &36.5          & 18.37        &  22.66          \\
Large Noise            &\textbf{52.34}                          &\textbf{24.14}                 & \textbf{22.42  }                 & 52.46           & 24.30          & 22.8         & 55.24           & 23.14       &33.39           \\
\hline
\textbf{Density} \\
Gradient                  &37.13                     &10.66                  & 12.87                  &\textbf{15.38}          & \textbf{10.51}         &13.42         & 16.61          & 14.65     &\textbf{11.02  }               \\
Stripes                  &36.12                             & 10.29                & \textbf{11.73 }                & \textbf{13.19}        & \textbf{10.04}          & 11.74          & \textbf{14.5 }         &14.57       &12.47                \\
\hline
Average                  &30.23                          & \textbf{13.62}                   & \textbf{14.37}                & 22.90          & 13.70          & 14.56          & \textbf{23.91}        & 16.14    &16.89                    \\

\bottomrule
\end{tabular}
\end{table*}

\subsection{Evaluation}
As discussed above,  to compare our method to other deep learning methods and some classic geometric methods, we use angle difference between our predicted normal and ground truth via MSER metric. In the experiment, we compare the improvement of our method on a single scale by local plane feature constraints and also show excellent performance on multiple scales. 

Table \ref{Tab01} shows the comparison of unoriented normal estimation using single scale methods discussed above. In the top row of the table, we show the results for varying levels of noise, from zero noise to high noise. The two rows in the middle show the results for point clouds with a non-uniform sampling rate. In each of the categories, we show the average for all shapes in one category. The last row shows the global average error over all shapes. As shown in Table \ref{Tab01}, we slightly improve on the state of the art on most single scale levels with the different noise levels and varying densities. As shown in Table \ref{Tab01}, with the noise added, our method can improve the normal results obviously. The reason for the promotion is that the plane-based classify can improve our network to extract more robust features that are more correlated with the local normal and can better remove the meaningless disturbance produced by the noise. In Table\ref{Tab01}, we can see when the patch scale is selected by 0.03 and 0.05, our method obtains slightly better results. In addition, in the case of less noise or no noise, we can also achieve competitive results compared to the method of PCPNet. The second and third row of Figure \ref{FIG:05} also show the improvements compared to single scale results of PCPNet\cite{guerrero2018pcpnet}. 
 
 \begin{figure}
  \centering
    \includegraphics[scale=0.3]{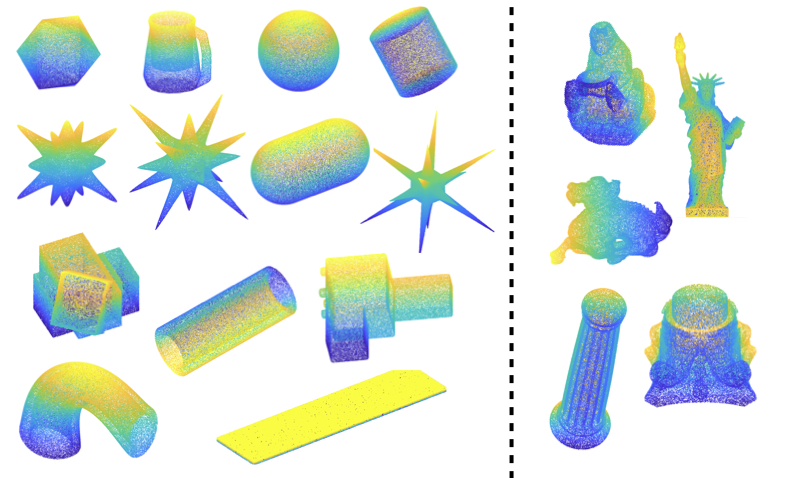}
  \caption{The divided test dataset. The shapes are divided into two simple man-made objects and complex organisms.}
  \label{FIG:09}
\end{figure}
 
Besides, we remove the organism objects from the test data of the PCPNet dataset shown in Figure \ref{FIG:09}. Here, we remove five models (right models).  We find that our method can improve the normal estimation of  man-made objects, Table \ref{Tab02} shows the results without the organisms, and on each level of noise level, our method is always can improve the final estimation on each scale. We can give more robust results compared to simple regress the patch normals. The point-wise feature constraints really can improve the results via extract more robust features. Our method is better for preserving plane information, especially when models have higher noise. But as shown in Table \ref{Tab01}, on the whole dataset, our method has no obvious performance on small scale (0.01), it is the possibility that false plane is rare in small scale and our method could give more improvements for large scale sampled patch.
 
 \begin{table}
\centering
\caption{Comparison of the RMS angle error for normal estimation of our single scale (0.03 and 0.05 respectively) results and PCPNet\cite{guerrero2018pcpnet} on the test dataset without the organism models. Our method can improve the accuracy of rigid models that have more regular planes. }
\label{Tab02}
\setlength{\tabcolsep}{2.5mm}{
\begin{tabular}{lcccccc}
\toprule
\multirow{2}{*}{} & \multicolumn{2}{c}{\textbf{Our Method}} & \multicolumn{2}{c}{\textbf{PCPNet\cite{guerrero2018pcpnet}}} \\
\cmidrule(r){2-3} \cmidrule(r){4-5}
&  $0.03$   &   $0.05$
&  $0.03$   &   $0.05$\\
\midrule
No Noise  &\textbf{4.76}    &\textbf{5.59}    &5.12   &5.82   \\
\hline
\textbf{Noise} \\
Small Noise   &\textbf{6.31}                      & \textbf{6.73}                     & 6.64                      &7.07                     \\
Middle Noise           &\textbf{11.65}                        & \textbf{11.17}                   & 12.14                       & 12.24                \\
Large Noise            &\textbf{18.66}                         &\textbf{16.31}                 & 19.02                 &16.96           \\
\hline
\hline
Average                  &\textbf{10.35}                        & \textbf{9.95}                   &10.73                      &  10.52                        \\

\bottomrule
\end{tabular}}

\end{table}

\begin{figure*}[h]
  \centering
    \includegraphics[scale=0.5]{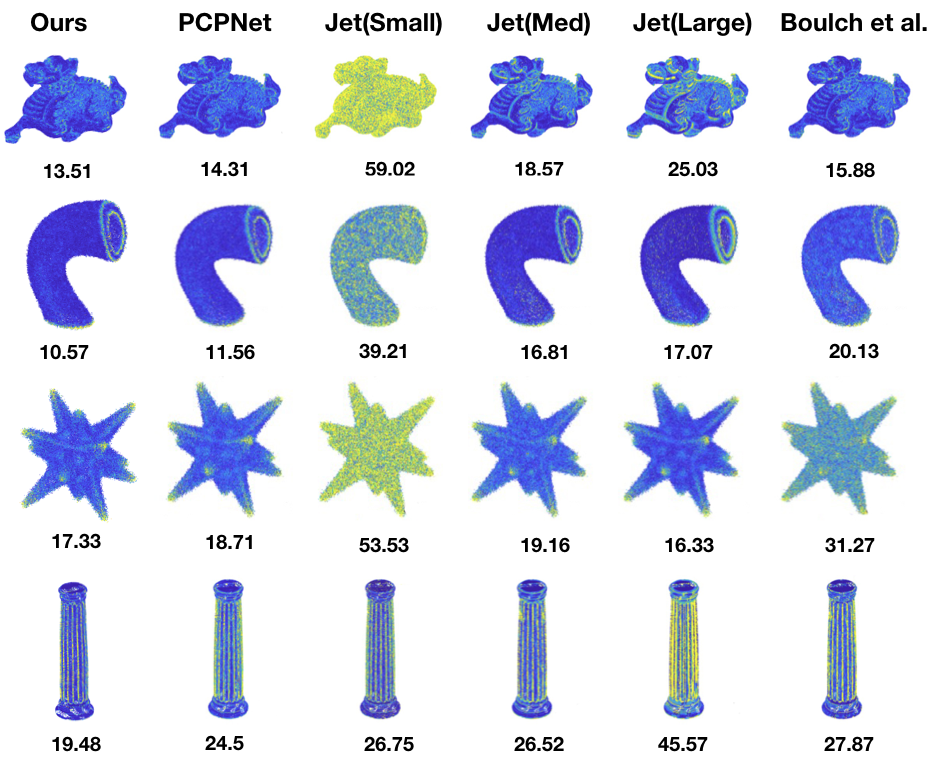}
  \caption{Qualitative comparison of the normal estimated by our method to other methods.  The point colors  correspond to the angular difference from the truth normals, mapped to a heatmap ranging from $0$ to $60$ degrees (blue to yellow). The numbers show the RMS error of each shape.}
  \label{FIG:04}
\end{figure*}

Then, we also show the outperformed results of multi-scale normals estimation. Show in Table \ref{Tab03}. We can observe the following general trends in these results: first, note that all of our methods consistently outperform competing techniques across all noise levels and compared to single scale results, the multi-scale selection methods can improve our results based on any single scale shown in Table\ref{Tab01}. It can be seen that our method outperforms almost other methods across all noise levels and most density variations. Especially compared to the PCPNet\cite{guerrero2018pcpnet}, which is the based network we used,  our approach can be greatly improved the normal estimation results compared to simple concatenation strategy used in PCPNet\cite{guerrero2018pcpnet}. Qualitative comparisons of the normal error on four shapes of our dataset are shown in Figure\ref{FIG:04}. Note that for classical surface fitting, small patch sizes work well on detailed structures like in the bottom row, but fail on noisy point clouds, while large patches are more tolerant to noise, but smooth out surface detail. Our method can obtain more consistent results compared to classic methods. In addition, compared to PCPNet\cite{guerrero2018pcpnet}, our multi-scale normal estimation method performs well. Show in Figure \ref{FIG:04}.

\begin{table*}[]
\centering
\caption{Final results for unoriented normal estimation. Shown are normal estimation errors in angle RMSE. For PCA\cite{hoppe1992surface} and Jet\cite{cazals2005estimating}, optimal neighborhood size for average error is chosen. For our approach, we display results via our scale selection use the multi-scale network. Our methods can improve the results for most noise levels. Here, multi1: three scales are 0.01, 0.03 and 0.05, multi2: three scales are 0.03, 0.05 and 0.07.}
\label{Tab03}
\begin{tabular}{lcccccccc}
  &\textbf{Ours(multi1)}     &\textbf{Ours(multi2)} &\textbf{Nesti-Net\cite{ben2019nesti}}  &\textbf{PCPNet\cite{guerrero2018pcpnet}}   &\textbf{HoughCNN\cite{Boulch2016Deep}}   &\textbf{PCA\cite{hoppe1992surface} }   &\textbf{Jet\cite{cazals2005estimating}}  \\
\hline
No Noise    & 7.30    &8.42   &\textbf{6.99}      &9.68                & 10.23        & 12.29     & 12.23 \\
Small Noise   &10.14      & 10.28       &\textbf{10.11}   &11.46                  & 11.62                           &12.87      &12.84   \\
Middle Noise           &17.54   &\textbf{17.49}        &17.63   &18.26                    & 22.66                       & 18.38                  &18.33\\
Large Noise            &22.43           & \textbf{22.23}    &22.28             &22.8                  & 33.39                     &27.5       &27.68  \\
Gradient     &\textbf{8.80} &10.36  &9.00     &11.74    &12.47      &13.66    &13.39\\
Stripes       & \textbf{8.37} &9.83   &8.47     &13.42    &11.02      &12.81    &13.13\\
\hline
Average                  &12.43       &13.10    &\textbf{12.41}               &  14.56                 & 16.9                          & 16.25    &  16.29    \\
\end{tabular}
\end{table*}

 \begin{figure*}[h]
  \centering
    \includegraphics[scale=0.47]{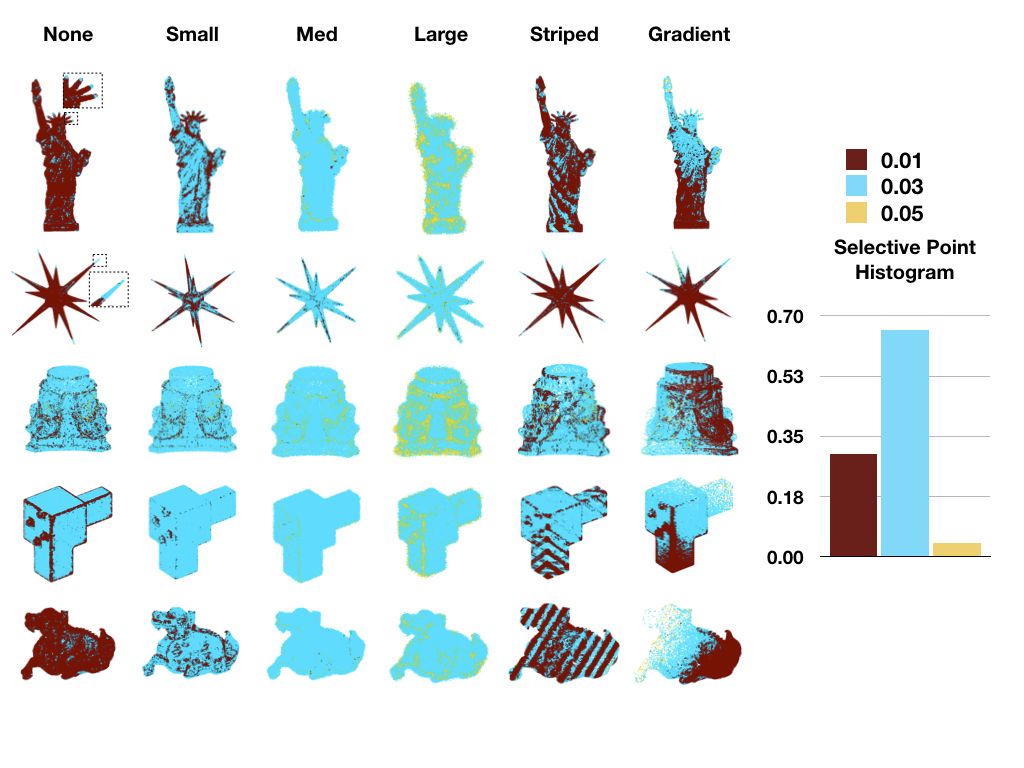}
  \caption{The visualization of the selected scale used in our network. The colors of the points correspond to the label of the scales (0.01, 0.03 and 0.05). On the right: we give the color coding and the distribution of point scale selection on the test dataset.}
  \label{FIG:06}
\end{figure*}

 \begin{figure*}[h]
  \centering
    \includegraphics[scale=0.47]{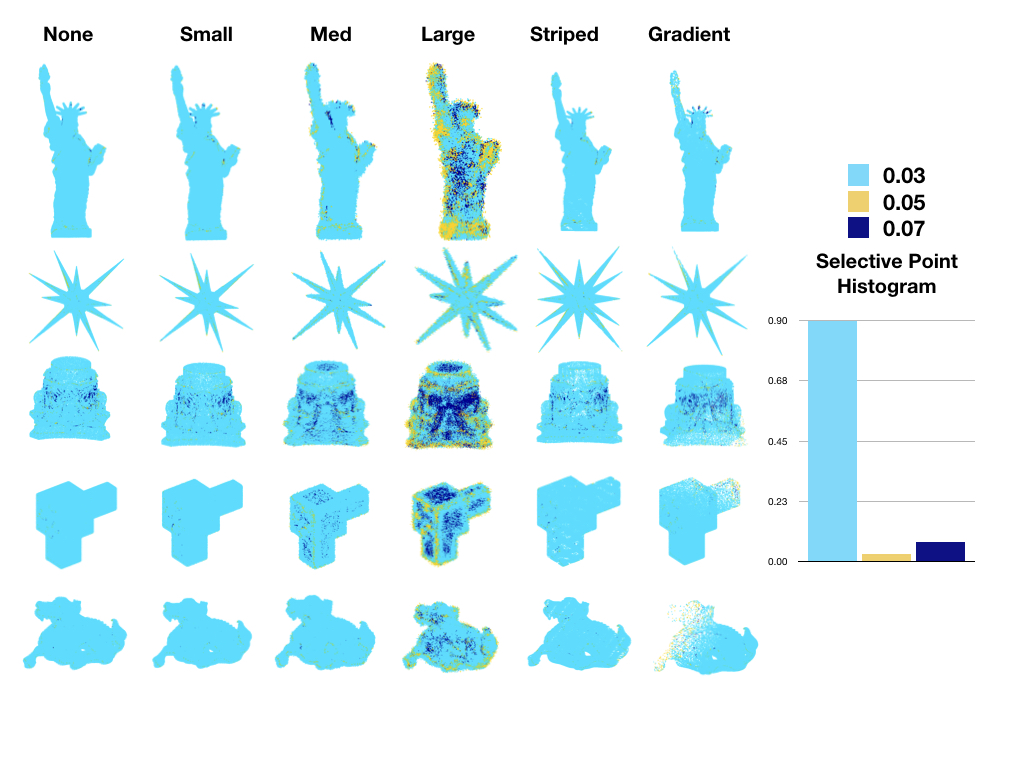}
  \caption{The visualization of the selected scale used in our network. The colors of the points correspond to the label of the scales (0.03, 0.05 and 0.07). On the right: we give the color coding and the distribution of point scale selection on the test dataset.}
  \label{FIG:07}
\end{figure*}

\subsection{Plane Constraints Results}
In our experiments, we also show the results of our local plane classification. As shown in Figure \ref{FIG:03}, from the top to down, we show the results for varying levels of noise, from zero noise to high noise. Our network can accurately distinguish between the set of points on the flat plane where the center point is and other sets of points. The robust results are achieved with different noise levels. In addition, we can see when a patch has a sharp edge, our network can divide the set of points into two categories and shows a clear boundary. This constraint can really improve the stage of feature learning of the input patch and we can obtain a plane-aware feature to estimate the normal of any scale patch. Figure \ref{FIG:05} shows that our LPFC can improve the single scale results both in 0.03 and 0.05 scale on the left test models shown in Figure \ref{FIG:09}. But as shown in Table \ref{Tab02}, On the whole dataset, our method has no obvious performance on small scale (0.03), it is the possibility that error part points are rare in small scale.

\subsection{Scale selection performance}
As shown in Table \ref{Tab01}, with noise increasing,  our methods can improve the normal estimation on the single scale architecture. The final evaluation results from different noise levels are affected by the choice of different scales. This means that the choice of scale has an important impact on the normal estimation. Besides, the first row of Figure \ref{FIG:05} also shows that our scale selection method can greatly improve the estimated results compared to the single scale results as shown in the second and fourth row of Figure \ref{FIG:05}. The final errors  in Table \ref{Tab03} also show that the scale selection strategy can greatly improve our results. In this section, we would like to show how the scales affect the final estimated normal. Shown in Figure \ref{FIG:06}, from left to right, the labels of scale selected in different noise levels are given. Three colors (blue, yellow and red) correspond to the scales respectively.  First, our multi-scale network tends to choose small-scale results in the case of low-level noise, because the point network can obtain accurate information from  small local patches which may include more relevant information.  When the part has rich local details, medium and small scale results are selected via our network, as shown in the example marked by dotted lines (1-2 row in the first column).  For the boundary part, small scale is generally chosen, the reason is that the small neighbors can better judge the surface where the current center point is. Show in 3-4 rows of Figure \ref{FIG:06}. Then, with the noise increasing, our network tends to select the medium and large scales to ensure that more information can be obtained to avoid that the errors (noise and outliers) are introduced. The first two columns of Table \ref{Tab03} also give the consistent result that large scale sampling can improve the normal estimation when the noise is increased in models.  As shown in Table \ref{Tab03}, when the multi scales are chosen from 0.03, 0.05 and 0.07, the normals estimated at high noise level will be slightly improved via multi-scale selection method. Furthermore, we compared the visual results from different multi-scale combinations. As shown in Figure \ref{FIG:07}, we use the scales 0.03, 0.05 and 0.07 as  the input of our networks. Compared to \ref{FIG:06}, When the minimum scale (0.03) is removed, the network loses the ability to select on the small scales, thus weakening its performance on the small noise models. However, the combination of larger-scales selection has a better performance in the large noise model due to the introduction of a larger scale. The same conclusion is also given in the table \ref{Tab03}. Finally, our proposed multi-scale network would like to choose a larger scale in the sparse region as shown in the last column of Figure \ref{FIG:06}. 

Besides, the distribution of points scale selection is also given using a histogram. Show in Figure \ref{FIG:06} and \ref{FIG:07}.  It can be seen from the histogram distributions of the selected points that for the PCPNet database, making more choices near the small scales is beneficial to the final estimation result. However, due to the limitation of our equipment, the comparison of more scales selection results was not given.

\section{Conclusion}
In this work, we propose a normal estimation method considered to use local plane features constraint strategy and select the best one from multi-scale results. We can improve the single scale results through the plane feature constraint mechanism, especially when the input of the network is a large scale patch. The local plane features constraint strategy can reduce the error caused by sampling error plane points. Besides, the multi-scale selection network enables the prediction of an optimal local scale and our experiments have analyzed the relationship between scale selection and point cloud normal estimation. Noise and local structure may both affect the best scale for normal estimation. The proposed method achieves state-of-the-art results relative to all other methods and demonstrates robustness to noise.

\section*{Acknowledgment}
We would like to thank all the reviewers for their valuable comments and feedback. This work was supported by NSFC (Nos. 61702079),  NSFC(Nos. 61632006) and NSFC(Nos. 61562062).

\section*{References}

\bibliography{mybibfile}

\end{document}